\documentclass[aps,prl,preprint]{revtex4}
\usepackage{graphicx}

\begin{document}

\title{GENEALOGICAL TREES FROM GENETIC DISTANCES}

\author{Luce Prignano} 
\address{Dipartimento di Fisica,
Universit\`a di Roma "La Sapienza",
I-00185 Roma, Italy}
\address{}
\author{Maurizio Serva} 
\address{Dipartimento di Matematica,
Universit\`a dell'Aquila,
I-67010 L'Aquila, Italy}

\bigskip

\date{\today}

\begin{abstract}

In a population with haploid reproduction any 
individual has a single parent in the previous generation. 
If all genealogical distances among pairs
of individuals (generations from the 
closest common ancestor) are known it is possible to exactly reconstruct their genealogical tree.
Unfortunately, in most cases, genealogical distances
are unknown and only genetic distances are available.
The genetic distance between two individuals
is measurable from  differences in mtDNA
(mitochondrial DNA) in the case of humans
or other complex organisms 
while an analogous distance can be also given
for languages where it is measured 
from lexical differences. 
Assuming a constant rate of mutation, these 
genetic distances are random and proportional 
only on average to genealogical ones.  
The reconstruction of the genealogical tree
from the available genetic distances
is forceful imprecise.
In this paper we try to quantify the error
one may commit in the reconstruction of
the tree for different degrees of randomness.
The errors may concern both topology of the tree
(the branching hierarchy) and, in case of correct topology,
the proportions of the tree (length of various branches).

\bigskip

\noindent
Pacs: 

\noindent
05.40.-a \, -Fluctuation phenomena, random processes, noise, and Brownian motion,

\noindent
87.23.Ge \, -Dynamics of social systems,

\noindent
87.23.Kg \, -Dynamics of evolution,

\noindent
89.75.Hc \, -Networks and genealogical trees.

\end{abstract}

\maketitle
 
\section{1. Introduction}

Haploid  reproduction implies that any individual has a single
parent in the previous generation.
Since some of the individuals may have the same parent, 
the number of ancestors of the present population 
decreases going backwards in time until a
complete coalescence to a single 
ancestor~\cite{King1, Kgen, Tavare}. 
Therefore, it is possible to construct a genealogical tree
whose various branching events connect
all the individuals living in the present time
to the single founder ancestor.
The genealogical distance between two individuals
is simply the time from their last common ancestor
and it may assume the maximal value
only when the common ancestor coincide with the founder.\\
In the limit of infinite population size,
most of the quantities remain random,
for example, this is the case of the probability density 
of genealogical distances in a single population.  
In fact, even in the thermodynamic limit,
this quantity varies for different populations
or, at different times, for the same population. 
The discovery of this non self-averaging behavior is due to the pioneering work of Derrida, Bessis, Jung-Muller and Peliti~\cite{DB,JM,Derrida}.

Nevertheless, if all genealogical distances among pairs
of individuals are known it is possible to exactly 
reconstruct their genealogical tree.
Unfortunately, in practice, genealogical distances
are unknown unless 
one has the relatives historical records
which is not the case of living organisms populations 
and of most of the linguistic groups
(Latin languages are an exception).
In most cases, only genetic distances are available.
These distances, in the case of humans
or other complex organisms, can be measured from the difference in mtDNA (mitochondrial DNA), which is inherited only from the mother
and, therefore, it undergoes to haploid reproduction~\cite{Krings1997,Krings2000}. In the case of languages, instead,
they are deduced from lexical differences~\cite{SP,SP2,PPS}. 
Assuming a constant rate of mutation, these 
genetic distances are random and they are
proportional only on average to genealogical distances.  
The reconstruction of the genealogical tree
from the available genetic distances
is forceful imprecise.
In this paper we try to quantify the error
one may commit in the reconstruction of
the tree for different degrees of randomness.
The errors may concern both topology of the tree
(the branching hierarchy) and, in case of correct topology,
the proportions of the tree (length of various branches).
The paper is organized as follows: sections 2 is devoted to
the deterministic process which is associated to
the genealogical distances. We also show there how to 
exactly reconstruct the genealogical tree form them.
In section 3 we define and discuss the random process associated to the genetic distances.
In section 4 we introduce a measure of topological distance
between two tree and we quantify how
topologically wrong is the tree
reconstructed from genetic distances.
In sections 5 we quantify the error
concerning the length of various branches of
topologically correct trees. 
Finally section 6 contains conclusions and outlook.
The paper is completed by an appendix where we have 
moved some lengthy calculations.

\section{2. Dynamics of genealogical distances\\ and trees reconstruction}

We consider a very general model of a population of 
constant size $N$ whose generations are not 
overlapping in time: any generation is replaced by a 
new one and any individual has a single parent. 
The stochastic rule which assigns the number of 
offspring to any individual can be chosen 
in many ways. In fact, for large population size, results 
do not depend on the details of this rule,
the only requirement is that the probability of having the same
parent for two individuals must be of order $1/N$ for large $N$.
Here we choose the Wright-Fisher rule:
any individual in the new generation chooses one parent 
at random in the previous one, independently on the 
choice of the others.

The genealogical distance between two given individuals 
is the number of generations from the closest common ancestor. 
For large $N$ distances are proportional to $N$, it is then useful to re-scale them dividing by $N$.\\
So let us define $d(\alpha,\beta)$ as the rescaled genealogical distance
between individuals $\alpha$ and $\beta$ in a population of size  $N$.
For two distinct individuals $\alpha$ and $\beta$
in the same generation one has
 
\begin{equation}
d(\alpha,\beta)= d(g(\alpha),g(\beta)) +\frac{1}{N}\;\;,
\label{dynamics}
\end{equation}
where $g(\alpha)$ and $g(\beta)$ are the parents
of $\alpha$ and $\beta$ respectively.
Accordingly with the Wright-Fisher rule, parents are chosen among all possible ones
with equal probability $1/N$ and, therefore,
$g(\alpha)$ and $g(\beta)$ coincide with probability $1/N$.
In this case the distance $d(g(\alpha),g(\beta))$ 
vanishes.
On the contrary, the parents of $\alpha$ and $\beta$ 
are distinct individuals $\alpha'$ and $\beta'$  
with probability $(N-1)/N$.
The above equation,
when considered all the $N(N-1)/2$ pairs,
entirely defines the dynamics of
the population and simply states that the
rescaled distance in the new generation increases by $1/N$
with respect to the parents distance.
Briefly, $d(\alpha,\beta)= 1/N$ with probability $1/N$
and $d(\alpha,\beta)= d(\alpha', \beta') +1/N$  with 
probability $(N-1)/N$.

This equation can be iterated for any of
the possible $N(N-1)/2$ initial pairs $\alpha$ and $\beta$, which correspond to the entries of an upper triangular matrix. 
The iteration stops when there is a coincidence of parents and
in this way all the distances $d(\alpha,\beta)$ can be calculated.

Shortly: iteration of equation \ref{dynamics} gives as output
the realization of the random $N(N-1)/2$ distances 
$d(\alpha,\beta)$ which are the entries of an upper 
triangular matrix containing
all the necessary information  for the reconstruction 
of the genealogical tree of the population.
The tree is completely identified by its topology
and by the time separation of all branching events.
There exist many methods that can be used for
this reconstruction, a simple one is the 
Unweighted Pair Group Method Average (UPGMA)~\cite{UPGMA}.
This algorithm works as follows: 
it first identifies the two individuals with shortest distance,
and put their branching at their time separation. 
Then, it treats this pair as a new single object whose distance
from the other individuals is the average of the distance of 
its two components.
Subsequently, among the new group of objects it identifies the pair with 
the shortest distance, and so on. 
At the end, one is left with only two objects
which represents the two main branches, whose distance gives the time position of the root of the tree.
Then, the time from the last common ancestor 
of all individuals in the populations results fixed.
 
This method works for any kind of upper 
triangular matrix representing distances among pairs of individuals,
not necessarily originated by the coalescent process.
In the coalescent case, nevertheless,
the method gives the correct tree
reproducing the historical branching events
and the correct time separations among them.
Notice, that at any time it chooses two individuals with shortest 
distance.  Then, it is easy
to realize that for the coalescent
the distance of the two individuals from any third one is the same. 
Therefore, in this case, all UPGMA averages are
between pairs with identical distances so that
also the resulting new common distances are the same.
 
Genealogical trees are very complex objects and genealogical distances 
are distributed according to a probability density which remains random in the limit of large 
population~\cite{Serva2,S}. Anyway, the
mathematical theory of coalescent gives us the ability to deduce some 
important information about their statistical structure.
Consider a sample of $n$ individuals in a population of size $N$, where $N$ is very large with respect to $n$.
The probability that they all have different parents in the 
previous generation is $\prod_{k=0}^{n-1}\, 
(1-\frac{k}{N})$.
Therefore, the probability that their ancestors are still 
all different in 
a past time $t$ corresponding to $tN$ generations
is $[\prod_{k=0}^{n-1}\, 
(1-\frac{k}{N})]^{tN}$. 
If $N$ is large compared to $n$ 
this quantity is approximately $\exp(-c_n t)$
where  $c_n = \frac{n\,(n-1)}{2}$. 
The genealogical tree results from this rule: 
the average probability density for the time lag
for the coalescent event for $n$ individuals is 
$p_n(t)=c_n \exp(-c_n t)$~\cite{King1,Avi}.
 
Therefore, the tree starts at the root
with the two main branches,
then, the following branching event
is at a random time lag $t_2$ 
with probability density $p(t_2)=e^{-t_2}$
and after this time the tree has three branches
(see Fig. \ref{tree3}). 
The next branching is after a time lag $t_3$ 
with probability density is $p(t_3)=3 \,e^{-3t_3}$
and then the tree has four branches, and so on.

\section{3. Random coalescent process}

As already mentioned, in the coalescent model, genealogical distances measure the time 
from the last common ancestor of two individuals. Nevertheless, 
in almost all real situations, we have to deal with genetic 
distances reconstructed from directly measurable empirical quantities.

In case of complex organisms, mtDNA is inherited only from the mother
and, therefore, it undergoes to haploid reproduction,
so in this case genetic distances are proportional to the 
number of mutations  occurred in the compared 
mtDNA sequences
(see, for example, ~\cite{Krings1997,Krings2000}).  
Analogously languages can be considered 
as haploid individuals whose vocabulary changes  
accumulate in time, in this case
genetic distances can be evaluated by lexical distances~\cite{SP,SP2,PPS}.

In both cases, an individual 
randomly accumulates mutations at a constant rate
and the genetic distance of a pair of individuals is the sum of the mutations that they accumulated since 
their last common ancestor (rescaled by  $N$).
As a consequence,
genetic distances are proportional only on
average to genealogical ones. 
Therefore, we have to modify the deterministic  
equation (\ref{dynamics}) in order to take into account this
randomness.
We may assume that increments in the genetic distance
have the simple form
 
 \begin{equation}
h(\alpha,\beta)=h(g(\alpha),g(\beta))+\gamma_{\alpha}
+\gamma_{\beta}
\label{dynamics2}
\end{equation}
where $g(\alpha)$ and $g(\beta)$ are the parents of $\alpha$ and $\beta$ respectively, while $\gamma_{\alpha}$ 
and $\gamma_{\beta}$ are random variables associated
to the mutations of $\alpha$ and $\beta$ .
They are zero if the genome of the parent is transmitted  unaltered and a positive constant 
if a mutation occurs.
We assume that the probability is
$ 1-\frac{\mu}{2N}$ for zero and $\frac{\mu}{2N}$ for the positive constant
$\frac{1}{\mu}$.
In a compact form:

\begin{eqnarray}
\gamma_{\alpha}=\left\{\begin{array}{cc} 0 &
prob=1-\frac{\mu}{2N}\\\frac{1}{\mu} & 
prob=\frac{\mu}{2N}\end{array}\right.
\label{gamma}
\end{eqnarray}

This rule grants that genetic distances are equal 
to genealogical distances on average,
in fact, the expected value of the
sum $\gamma_{\alpha} +\gamma_{\beta}$ is $1/N$.

Notice that we compare genealogical distances 
generated by (\ref{dynamics})
with genetic ones generated by (\ref{dynamics2}).
Since they describe two aspects of the same population,
the family history must be the same.
This means that the realization 
of the part of the process 
which assigns parents in (\ref{dynamics})
and in (\ref{dynamics2}) must also be the same ($\alpha\to
g(\alpha)$), 
the only difference lays in the deterministic/random 
nature of the distance increment. In other words, 
the parents of two individuals $\alpha$ and $\beta$ 
are unequivocally assigned. 
Then, their genealogical distance is
$d(\alpha,\beta)=d(g(\alpha),g(\beta))+1/N$, while 
the genetic one is $h(\alpha,\beta)=h(g(\alpha),g(\beta))+
\gamma_{\alpha}+\gamma_{\beta}$
where $\gamma_{\alpha}$ and $\gamma_{\beta}$ are the previously defined random variables.

\begin{figure}[htbp]
\begin{center}
\includegraphics[width=13cm]{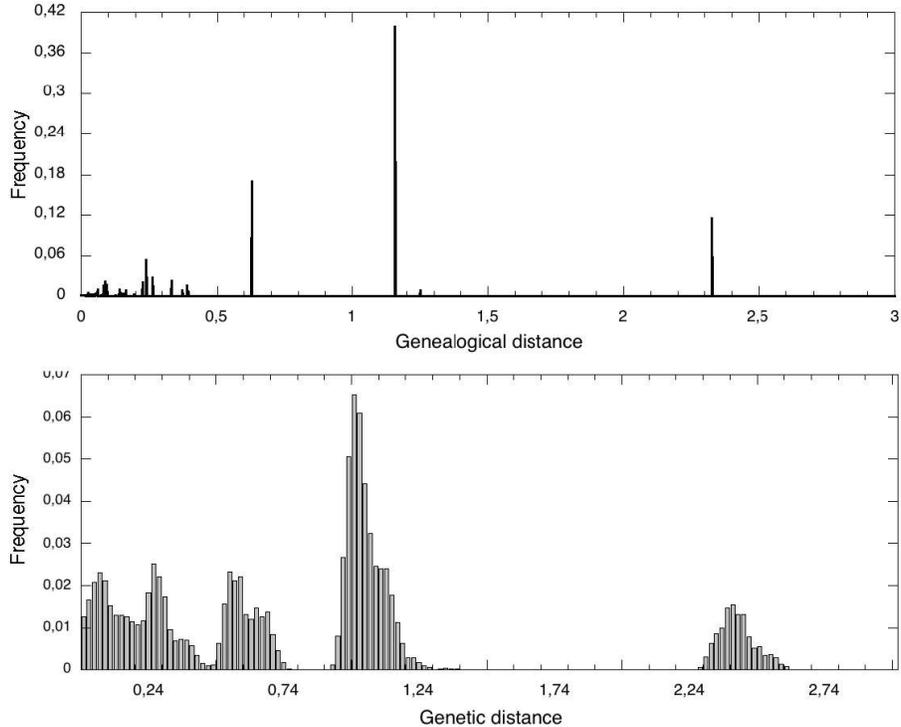}
\caption{Frequencies of the genealogical
and genetic distances ($\mu$=50) in a population
of 700 individuals.
The realization of the genealogical
process (the family history) is the same for the two distributions.}
\label{spikes}
\end{center}
\end{figure}

In order to have a qualitative idea of the differences
between the set of genetic distances and the set of genealogical ones we plot in Fig.\,\ref{spikes} \
the frequency of the distances of the two sets.
We have used the same realization of the
process for 700 individuals
and we have chosen $\mu= 50$ for the genetic distances.
We can see that while genealogical distances
may assume only few values where their distribution
has spikes ~\cite{Serva2,S}, 
the genetic ones are dispersed around them.
The dispersion decreases when $\mu$ increases
and when $\mu= 2N$
genetic distances lose their randomness in mutations,
and they equal, for any realization, 
the associated genealogical ones ( compare
(\ref{dynamics}) with 
(\ref{dynamics2}) and (\ref{gamma}) ). 
In case of a very large population ($N \to \infty$)
the coincidence of the two sets 
of distances is recovered in the $\mu \to \infty$ limit.

\section{4. Wrong tree reconstruction: topology}

The problem, in most practical cases, is
that genealogical distances are unknown
and one would like to reconstruct the genealogical tree
of a population from the measured genetic distances.
This is the case of biology where
strands of mtDNA are compared as well of lexicostatistics
where vocabularies or grammar structures
take the same role of mtDNA.

As mentioned in previous section,
when $\mu = 2N$ equations (\ref{dynamics}) and
(\ref{dynamics2})
coincide and randomness in mutations is lost.
In this limiting case genetic and genealogical distances 
are equal and not only the frequency distributions
are identical, but also the family trees reconstructed by 
UPMGA will be exactly the same.
For smaller values of $\mu$, we expect 
that the fidelity level of reconstruction of a tree decreases.
Then, we would like to have a quantitative information 
on the difference between the trees reconstructed
from the matrices of genealogical and genetic distances. 

\begin{figure}[htbp]
\begin{center}
\includegraphics[width=16cm]{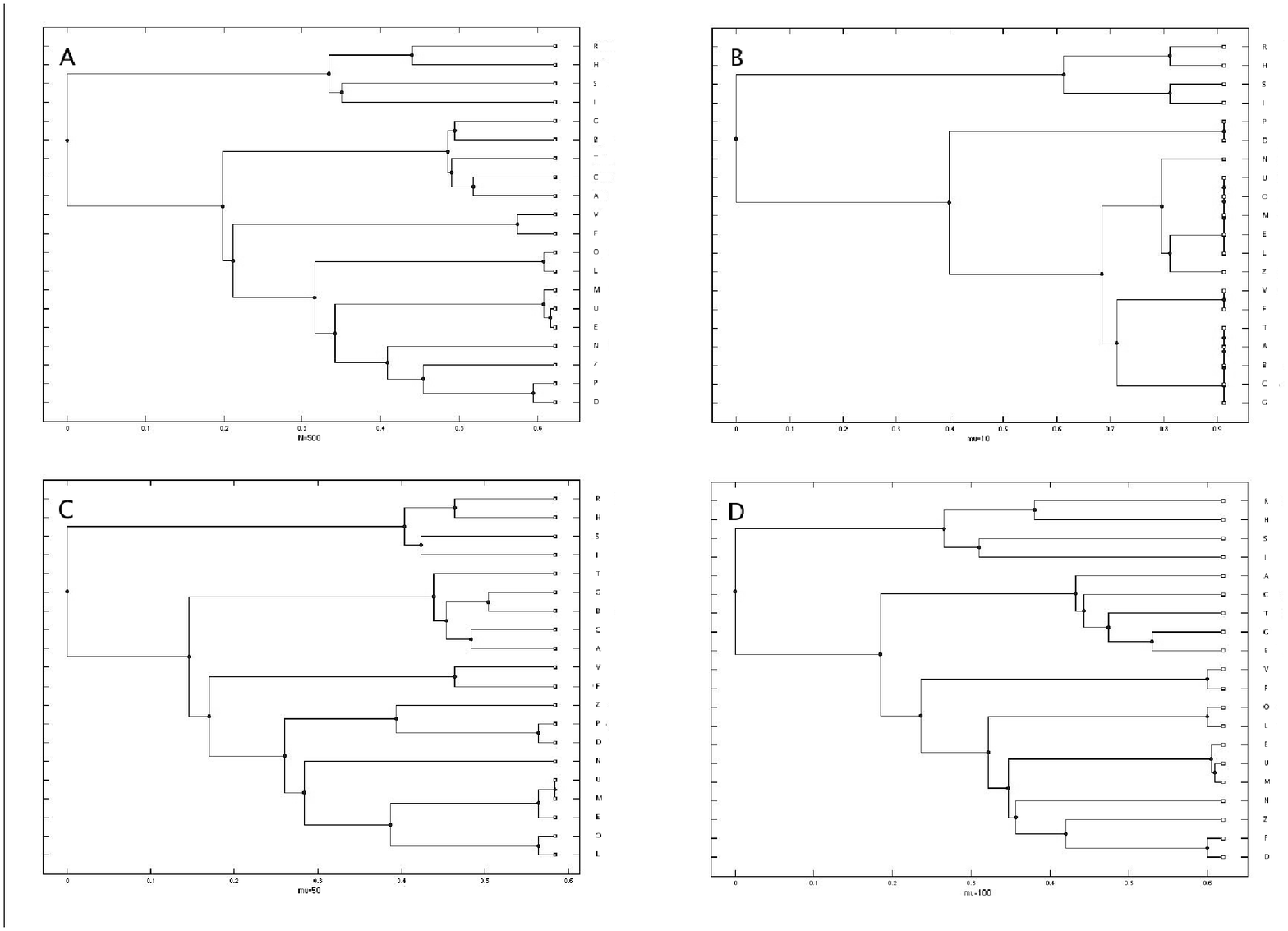}
\caption{Four trees of twenty leaves,
corresponding to a sample of $20$ individuals in a 
population of $500$.
The first tree is reconstructed from the genealogical 
distances and the others from the genetic ones 
($\mu$=10 for $B$, $\mu$=50 for $C$ and $\mu$=100 for $D$).
The quality of the reproductions of the first tree 
(the correct one) by the others is lower
for smaller $\mu$ both
for what concerns topology and time depths.}
\label{tree}
\end{center}
\end{figure}

A qualitative understanding of the problem
is immediate from Fig.\,\ref{tree} \ where four
trees of twenty leaves are reconstructed.
The first and correct one with label $A$, 
is associated to the genealogical 
distances and the remaining three to the genetic ones 
for three different values of $\mu$.
The realization of the genealogical
process (the family history) 
is the same for the four pictures.
One can see that the quality of the reconstruction
decreases for smaller $\mu$.
In fact, the tree with label $D$, which corresponds
to $\mu= 100$, is topologically quite correct,
with a couple of wrong clades (see leaves $G,B,T,C,A$
and $M,U,E$) at the lower level. 
Also the separation times of the branches are not
so different from the correct ones.
For $\mu= 50$ with label $C$ and $\mu= 10$ with label $B$
the quality of reproduction reduces,
clades are wrong also at a higher level
and times depths are quite different from the right ones.

We start the quantitative study of the quality 
of reconstruction considering the simplest situation 
of tree with three leaves.
The topology of a three leaves tree is completely determinate by the pair of individuals 
that first match together because 
their distance is the smallest.
Consequently, the genealogical and the genetic trees reconstructed by the UPGMA will
have the same topology if the same pair of individuals has 
both the smallest genetic and genealogical distance.
Let us call $\alpha$, $\beta$ and $\gamma$ the three individuals,
and assume that $\alpha$ and $\beta$ are the pair with the smallest 
genealogical distance $d(\alpha,\beta)$.
By the argument in Section 2 we know that $d(\alpha,\beta) = t_3$
and $d(\alpha,\gamma) \, = d(\beta,\gamma) = t_2+ t_3$, where
$t_2$ and $t_3$ are independent exponentially distributed variables with 
average $1$ and $1/3$ respectively.
Then let us consider the two following events concerning
genetic distances, the first that we call $A$ is
 
\begin{equation}
h(\alpha,\beta) < \min\{ h(\alpha,\gamma) ; \, h(\beta,\gamma) \} 
\label{minsep}
\end{equation}
If $A$ is satisfied, the topology of the genetic tree 
reconstructed by UPGMA is the correct one
since it is the same of that of the genealogical tree.
The second that we call $B$ is

\begin{eqnarray}
\begin{array}{cc} 
h(\alpha,\beta) = \min\{ h(\alpha,\gamma) ; \, h(\beta,\gamma) \} 
\\\
h(\alpha,\gamma) ; \, \neq \,h(\beta,\gamma)  
\end{array}
\label{esep}
\end{eqnarray}
which corresponds to an ambiguous (but unlikely) situation for UPGMA which
will be able to reconstruct correctly the tree with probability $1/2$.
The third that we call $C$ will be

\begin{equation}
h(\alpha,\beta) = h(\alpha,\gamma) =  \,h(\beta,\gamma)  
\label{esep2}
\end{equation}
which is also ambiguous (and even more unlikely). In this case, UPGMA 
will be able to reconstruct correctly the tree with probability $1/3$.

Let us now call $P(A \, | \, t_2 \, , t_3)$ the probability
of the event $A$ given the realized values  $t_2$ and $t_3$,
and $P(B \, | \, t_2 \, , t_3)$ and $P(C \, | \, t_2 \, , t_3)$ the
equivalent conditional probabilities for the events $B$ and $C$ respectively.
Let us also call $P(W \, | \, t_2 \, , t_3)$
the probability of a wrong reconstruction of the tree correspondingly
to $t_2$ and $t_3$. We have

\begin{equation}
P(W \, | \, t_2 \, , t_3) \, = \, 1 \, 
-\, P(A \, | \, t_2 \, , t_3) \, - \,
\frac{1}{2}P(B \, | \, t_2 \, , t_3) \, - \, 
\frac{1}{3}P(C \, | \, t_2 \, , t_3)  
\label{wrong}
\end{equation}

\begin{figure}[htbp]
\begin{center}
\includegraphics[width=8cm]{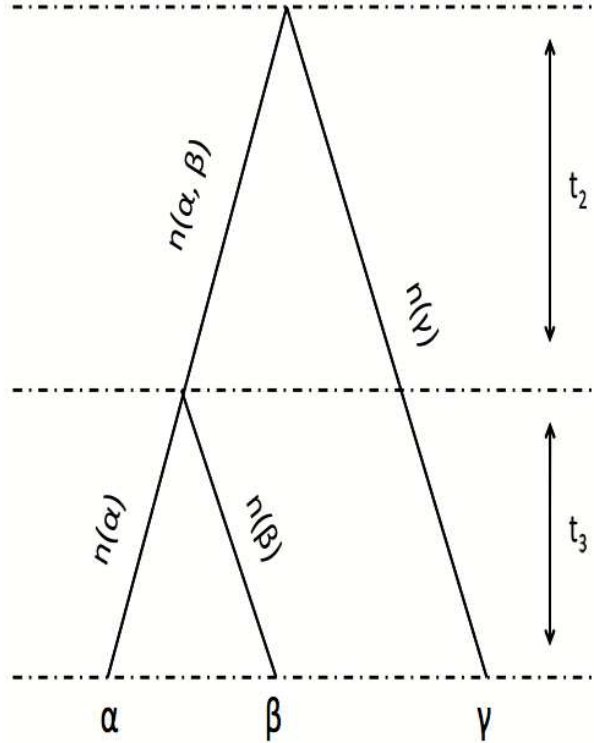}
\caption{Outline of a three leaves tree. $n(\alpha)$, $n(\beta)$, $n(\gamma)$ and $n(\alpha\beta)$ 
are the numbers of mutations divided by $\mu$.}
\label{tree3}
\end{center}
\end{figure}

Now we call $n(\alpha)$ the number of mutations along the branch $\alpha$
divided by $\mu$,
as shown in Fig. \ref{tree3}, analogously we define
$n(\beta)$,  $n(\gamma)$ and $n(\alpha\beta)$ as the numbers of
mutations divided by $\mu$ along the branches indicated in Fig. \ref{tree3}.

We will have  $h(\alpha,\beta) = n(\alpha)+n(\beta)$, 
$h(\alpha,\gamma) = n(\alpha)+n(\alpha\beta)+ n(\gamma)$
and $h(\beta,\gamma) = n(\beta)+n(\alpha\beta)+ n(\gamma)$.
The advantage is that the four new variables are 
independent and can be obtained as the sum of variables
of type (\ref{gamma}) where the sum goes on a number which
is $N$ times the time lag of the associated branch.
Namely, $t_3$ for $n(\alpha)$ and $n(\beta)$, 
$t_2$ for $n(\alpha\beta)$ and $t_2+t_3$ for $n(\gamma)$. 
 
\begin{figure}[htbp]
\begin{center}
\includegraphics[width=13cm]{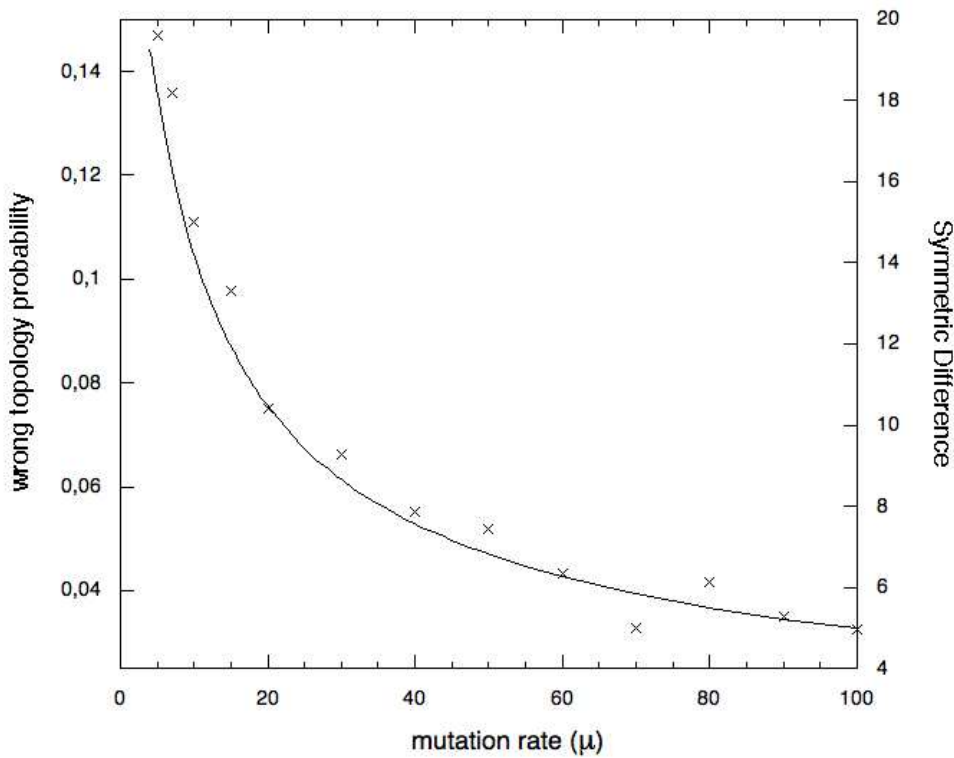}
\caption{Expected Symmetric Difference between 
a genealogical tree and the associated genetic 
tree plotted with respect $\mu$. 
The full line corresponds to the probability of a wrong
topology P(W) for a three leaves tree while crosses 
correspond to the ESD of a 20 leaves tree, estimated numerically. 
In the first case, computed exactly,
the ESD is twice the probability of a wrong topology.}
\label{td}
\end{center}
\end{figure}

Given this construction we can trivially but painfully
compute (see the Appendix) the conditional probability  $P(W \, | \, t_2 \, , \,t_3)$
and, then,  the absolute probability of a wrong tree  $P(W)$ 
as the marginal of the joint probability $P(W \, | \, t_2 \, , t_3)p(t_2)p(t_3)$
where $p(t_2)$ and  $p(t_3)$ are the exponential densities
previously described.
The probability of a wrong tree $P(W)$ is plotted in Fig. \ref{td} with respect to the parameter $\mu$
in the case of three individuals in
a large population ($N \to \infty $).

If we take into account more than three individuals the situation immediately becomes more 
complicated since the possible tree topologies increase exponentially with the number 
of leaves. So we need to introduce a measure
of difference between the genealogical tree and an associated genetic one.
The simplest tree distance measure is the Robinson-Foulds Symmetric Difference
\cite{RF}, which only depends on  the topology of the two tree and not on 
the differences in branches length.

The Symmetric Difference (SD) is computed by considering all possible branches 
that may exist in the two trees. 
Each inner branch, i.e. a branch connecting two nodes or one node to the root, 
identifies a clade in the set of leaves. The resulting distance is simply the number 
of clades present in one of the considered trees but not in the other. 
Therefore, two identical trees have zero SD, but it is 
sufficient to
exchange two leaves on one of them to have a non zero SD.

In general SD has not an immediate statistical interpretation, i.e.
we cannot say whether a larger distance is significantly larger than a smaller one.
Anyway, in the particular case of trees with only three leaves 
the expected symmetric distance is twice the wrong topology probability $P(W)$. 
In fact, in a three leaves tree there is only one clade and the Symmetric 
Distance is equal to 0 in the case of correct topology (if both trees have the same clade)
and is equal to 2 in the case of wrong one (if clades are
different). Consequently, in this simple case, 
the expected SD (that we call ESD) is given by the relation
  
$$\textrm{ESD}=2\cdot P(W)+0\cdot(1-P(W))=2\cdot P(W).$$
 
In order to compute numerically
the ESD between a genealogical tree 
and the associated genetic one with parameter $\mu $ we use
the following procedure:
we take 20 individuals in a population of 500 
(a large one) and we use UPGMA to reconstruct 
their genealogical tree from a realization of the genealogical distances matrix. 
Then, we construct several associated genetic trees ($5$ for $\mu<15$, $10$ for greater values)
and we compute their averaged SD 
with respect to the associated genealogical tree.
We start again with a new realization
of the matrix of genealogical distances and we repeat
the procedure, ending with a new averaged SD.
We do it many times (from $6$ for $\mu=5$ to $30$ for $\mu=100$) and, finally, we take the mean of 
all averaged SD and we end with a quantity that should
be very close to ESD.
The number of genealogical trees and that of the associated genetic trees that we use for estimating the ESD
increases with $\mu$ since we observe an increasing fluctuation in the SD values.

In Fig. \ref{td} we plot the estimated ESD of the 20 leaves tree. 
We also plot the exactly computed $P(W)$ of a three 
leaves tree which is one half its ESD. 
We find out, unexpectedly, 
that they only differ for a factor due to the total number of clades, which depends on the number of leaves.

\section{5. Wrong tree reconstruction: branches length}

Now we study the problem of differences among
branches length in genealogical and genetic trees.
We first derive the probability
of the error between the genetic 
and genealogical distances of two individuals, 
and then we 
will sketch out the computation of the errors for the distances in a tree of three individuals. 
We will consider only the cases in which
genetic and genealogical tree have the same topology. 
In this way the integral of the probability 
of the error (errors for the three leaves case)
will be equal to the total probability 
of having the right topology $P(R)=1-P(W)$. 
Obviously, the right topology condition is always 
satisfied for two individuals ($P(R)=1)$.

In a tree of two individuals we have only a single 
genealogical distance to be compared with a single 
genetic one.
Let us call $t$ the genealogical distance and $h$ 
the genetic one, that is the number of mutations 
on both the genealogical branches divided by $\mu$. 
We call $k=\mu \,h$ the total number of mutations 
on these two branches, then $t$ and $h$ 
will be equal only if $k=\mu t$. 
By using equations (\ref{gamma})
we can write the expression of
the conditional probability of having $k$ 
mutations along the two genealogical 
branches with lengths  $tN$ and, therefore,
total length $2tN$: 

\begin{equation}
p_\mu (k|2tN)=\left(\begin{array}{c}2tN 
\\k\end{array}\right)\left(\frac{\mu}{2N}\right)^k\left(1-\frac{\mu}{2N}\right)^{2tN-k}
\sim \frac{e^{-\mu t}(\mu t)^k}{k!}.
\label{kt}
\end{equation}
where the approximation holds for large $N$.

One easily gets the conditional 
averages  $<$$k$$>= \mu t $
and $<$$(\,k-<k>)^2$$> = \mu t $. 
It is than straightforward to define the  
error between genetic and genealogical distance 
as $\epsilon = \frac{h-t}{\sqrt{t}}$,
in this way, in fact, one gets the
conditional (with respect to t) averages
$<$$\epsilon$$>= 0$ and $<$$\epsilon^2$$> = 1/\mu $
which do not depend on $t$.
As a consequence of this independence
the absolute averages of $\epsilon$ and $\epsilon^2$
coincide with the conditional ones.
The conclusion is that the typical error in evaluating the
distance from the common ancestor
grows linearly with $1/\sqrt{\mu}$

The independence of the two conditional 
averages from $t$
does not implies that the conditional 
probability  density for $\epsilon$ is itself
independent on $t$. In fact, since one has
$h=k/\mu$ and, therefore,
$\epsilon = \frac{k-\mu t}{\mu \sqrt{t}}$,
the conditional probability  density
of the error turns out to be

\begin{equation}
p_{\mu}(\epsilon|t)=\sum_{k=0}^\infty
\, \frac{e^{-\mu t}(\mu t)^k}{k!} \;
\delta( \epsilon - \frac{k-\mu t}{\mu \sqrt{t}}).
\label{kkt}
\end{equation}
where the $\delta(\cdot)$ are Dirac delta functions.
This conditional density for $\epsilon$ given a 
genealogical distance $t$,
at variance with its two first moments,
clearly depends on $t$.

Finally, the absolute probability density
$p_\mu(\epsilon)$ can be calculated as the 
marginal of the joint 
probability density $p_{\mu}(\epsilon|t)p(t)$, 
where  $p(t)\,$$=$$\,\exp(-t)$  is the 
density of the genealogical distances. 
We obtain

\begin{equation}
p_{\mu}(\epsilon)=\sum_{k=0}^\infty
\, \frac{e^{-(\mu+1) l}(\mu l)^k}{k!} \;
\frac{2\mu l \sqrt{l}}{\mu l + k}.
\label{kktt}
\end{equation}
where we have to use for $l$
the following definition
\begin{equation}
\sqrt{l}= 
\frac{\sqrt{(\epsilon \mu)^2 +4 \mu k} -\epsilon \mu}{2 \mu} .
\label{eps3}
\end{equation}

\begin{figure}[htbp]
\begin{center}
\includegraphics[width=13cm]{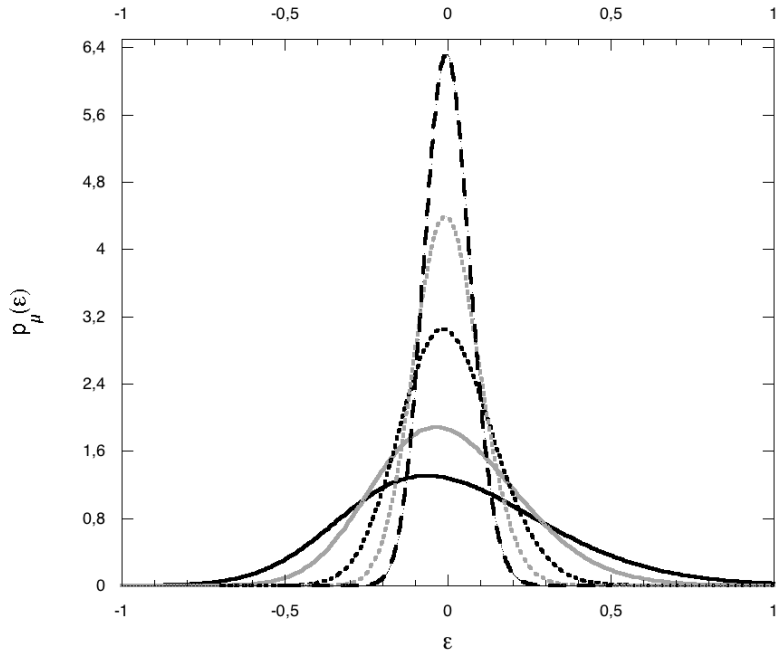}
\caption{The probability distribution of the error $p_\mu(\epsilon)$ 
is plotted for $\mu=10$, $\mu=20$, $\mu=50$, $\mu=100$ 
and $\mu=200$. The curves become sharper
for increasing values of $\mu$.}
\label{pe}
\end{center}
\end{figure}

In Fig.\,\ref{pe} \,  $p_\mu(\epsilon)$ is 
plotted for some values of $\mu$. 
In the limit $\mu$$\to$$\infty$ the distribution becomes a Dirac delta function centered in zero according with the fact
that the variance goes to zero as $1/\sqrt{\mu}$.
 
Now, let us come back to the case of a genealogical 
tree with three leaves. 
The study of this kind of structure, even if much simpler than the genealogical tree of an entire population, 
gives us important information since concerns the reconstruction of top of the tree. 
Indeed, to reconstruct in the correct way this part of the genealogical tree, 
that is the part going from the founder
to the three most recent ancestors 
of the entire population, 
means to rightly identify the three main sub-populations and their separation times.

We have seen the probability of having the correct topology for a three leaves tree from the genetic matrix 
by using UPGMA. In the following, we will give
a sketch of the derivation of the errors distribution
of its two characteristic distances.
Hereafter we restrict our analysis to the non-ambiguous situation in
which the topology of the genetic tree is 
the same as that of the genealogical one. 
We will refer to the scheme in Fig.\,\ref{tree3} and to the notation used in Section\,4. 

The genetic tree can be characterized by two distances: 
$h\, = \, h(\alpha,\beta)$, separating the
individuals with minimal genealogical distance $t_3$ and 
$H=\left[h(\alpha,\gamma)+h(\beta,\gamma)\right]/2$, calculated by
UPGMA as the mean value of the major distances and therefore separating the third individual from the others. 
The distance $H$ does not correspond to the maximum genetic distance
but it is the mean value of the two major distances and gives a better estimate of the coalescence time 
$T=t_3+t_2$ of three individuals to the common ancestor.
 
We have seen that on average $h$ coincides with $t_3$, $H$ on average is equal to $T$, but, in general, 
the genetic distances will be different from the 
genealogical ones and the branches of the tree reconstructed from the genetic matrix by UPGMA 
will not have the same length as those of the genealogical tree.

Here we define, in analogy with the two leaves case,
the errors $\epsilon_1$ and $\epsilon_2$ of $h$ and $H$ 
by the equations:
\begin{eqnarray}
\begin{array}{cc} 
&\epsilon_1 \,= \, \frac{h-t_3}{\sqrt{t_3}};\\
&\epsilon_2 \,=\, \frac{H-T}{\sqrt{T}}     .
\end{array}
\label{errel}
\end{eqnarray}
The variables $\epsilon_1$ and  $\epsilon_2$ vanish when the genetic distances equal the genealogical ones. 

In order to compute the probability density of the errors
we use the relations introduced in previous section 
and we rewrite the relations (\ref{errel}) in the form

\begin{eqnarray}
\begin{array}{cc} 
&\epsilon_1 = \frac{n(\alpha)+n(\beta)-t_3}{\sqrt{t_3}};\\
&\epsilon_2 = \frac{n(\alpha)+n(\beta) + 2\, n(\alpha\beta)+
    2\,n(\gamma)-2T}{2 \,\sqrt{T}} .
\end{array}
\label{errel2}
\end{eqnarray}
Since $n(\alpha)$, $n(\beta)$ and $n(\gamma) \,
+ \, n(\alpha\beta)$
are independent variables which we described in previous section
it is straightforward but painful to compute the conditional probability 
$p_\mu=p_\mu(\epsilon_1,\epsilon_2 | t_2, t_3)$. 

The sketch goes as follows: first we write the joint conditional probability for the independent variables
$n(\alpha)$, $n(\beta)$ and $n(\gamma) \,
+ \, n(\alpha\beta)$ as

\begin{equation}
p_\mu(n(\alpha)|t_3) p_\mu(n(\beta)|t_3) 
p_\mu(n(\gamma) +n(\alpha\beta)|-2t_2-t_3)
\label{jjj}
\end{equation} 
where the explicit expression 
for the probabilities $p_\mu(n|t)$
is shown in the Appendix (see (\ref{pn})).

Then we are able to compute 
$p_\mu= p_\mu(\epsilon_1,\epsilon_2 | t_2, t_3)$ as the sum
$p_\mu=p_\mu(A)+\frac{1}{2}p_\mu(B)+\frac{1}{3}p_\mu(C)$
where $p_\mu(A)$ is obtained by the sum of the 
conditional probability (\ref{jjj})
over all the triplets $n(\alpha)$, $n(\beta)$
and $n(\gamma) + n(\alpha\beta)$
which satisfy condition A 
of section 4 and relations (\ref{errel2}).
Analogously, we can compute $p_{\mu}(B)$ 
and $p_\mu(C)$.
Once we have   $p_\mu(\epsilon_1,\epsilon_2 | t_2, t_3)$
we can compute the joint probability
\begin{equation}
p_\mu(\epsilon_1,\epsilon_2 , t_2, t_3)
=p_\mu(\epsilon_1,\epsilon_2 | t_2,t_3) 3 e^{t_2+3t_3}
\end{equation} 
and after integration over $t_2$ and $t_3$ we obtain the joint density
$p_\mu(\epsilon_1,\epsilon_2)$.
The normalization of this density equals the 
probability $P(R)$ of correct topology identification by UPGMA. 

More complex tree could be considered in principle,
both for what concerns topology and branches length,  
nevertheless, the number 
of calculations increases exponentially.

\section{6. Conclusions and outlook}

The inner randomness of genetic mutations is an obstacle 
for a safe reconstruction of a genealogical tree.
A wrong reconstruction is more probable the smaller is the 
probability of a mutation.
This is a serious problem since in many cases 
in biology the distances are measured by a molecular
clock which is obtained comparing short strands of
DNA which slowly accumulate errors along reproduction events
(see, for example,~\cite{Krings1997,Krings2000}).

We are able to quantify, in simple but relevant cases, 
the probability of a wrong reconstruction of a tree, 
both for what concerns the topology and the proportions.
We can, for example, give the error concerning the time separation of two species using results in section 5 
and we are also able to decide the probability of a wrong reconstruction for a family tree of three species 
using results in section 4.

We plan to continue this investigation in order to better quantify the difference between the genetic and 
the genealogical matrices of distances.
We think for example to the possibility of introducing a measure of distance between the associated probabilities of distances.
Even more important, it is to find 
a method to estimate the value of the parameter 
$\mu$ given an empirical distribution 
of genetic distances.
This would be useful for situations in which
the value of $\mu$ is not known $a \,\,priori$ as,
for example, for the languages in the Indo-European and
Austronesian groups~\cite{SP,PPS,SP2}. 
This method would allow us to use the results 
of this paper for evaluating the fidelity level of phylogenetic trees reconstructed from empirical lexical distances.

\smallskip

\section{Acknowledgments}
We thank Filippo Petroni for many discussion concerning
both the ideas contained in this paper and their
numerical implementation.

\appendix
\section{APPENDIX: Wrong Topology Probability P(W)}
\label{AA}

In this appendix we will use the notation introduced 
in Section 3 and shown in Fig.~\ref{tree3}.   
We have that the probability $P(R)$ of having the
correct genealogical tree from the genetic distances matrix 
is:
\begin{equation}
P(R)=P(A)+\frac{1}{2}P(B)+\frac{1}{3}P(C)\,.
\end{equation}

$P(A)$ is the probability of event $A$, i.e., 
the probability of having genetic distances among 
individuals $\alpha\,,\beta$ and $\gamma$ satisfying 
the inequality (\ref{minsep}). Using the independent 
variables (\ref{minsep}) rewrites as

\begin{equation}
\max\{ n(\alpha) ; \, n(\beta)\} < n(\alpha\beta)+ n(\gamma)\,; 
\label{nminsep}
\end{equation}

$P(B)$ is the probability of event $B$, while $P(C)$ is the one of event $C$. 
The events $B$ and $C$ occur respectively if the genetic distances satisfy the conditions  (\ref{esep}) and (\ref{esep2}) which rewrite respectively as
\begin{eqnarray}
\begin{array}{cc} 
\max\{ n(\alpha) ; \, n(\beta)\} = n(\alpha\beta)+ n(\gamma)\,; 
\\\
n(\alpha) \, \neq \,n(\beta)\,.  
\end{array}
\label{nesep}
\end{eqnarray}
and
\begin{equation}
n(\alpha) \, = \,n(\beta) =  n(\alpha\beta)+ n(\gamma)\,.  
\label{nesep2}
\end{equation}

Let us define $n_1$ the maximum between $n(\alpha)$ and $n(\beta)$, $n_2$ the minimum ($n_1\,$$\geq$$\,n_2$), 
and $n_3=n(\alpha\beta)+n(\gamma)$.
Then, the probability $P(A)$ corresponds to the probability of 
having $n_1\,$$\geq$$\, n_2$ and $n_1\,$$<$$\,n_3$. 
Therefore, we have to determine the total probability for the set of triplets $\{n_1\,;\,n_2\,;\,n_3\}$, where 
the variables $n_i$ can take 
only values multiple of $1/\mu$ and have to satisfy the conditions  $n_1\,$$\in$$\,[0,\infty)$\,,
$n_2\,$$\in$$\,[0,n_1]$ and $n_3\,$$\in$$\,(n_1,\infty)$.

If we consider separately the cases $n_1\,$$<$$\, n_2$ and $n_1\,$$=$$\,n_2$ we immediately obtain:
\begin{equation}
P(A)=2\sum_{n_1=1/\mu}^{\infty}\sum_{n_2=0}^{n_1-1/\mu}\sum_{n_3=n_1+1}^{\infty}p_{\mu}(n_1,n_2,n_3)+
\sum_{n_1=0}^{\infty}\sum_{n_3=n_1+1/\mu}^{\infty}p_{\mu}(n_1,n_1,n_3)\,,
\end{equation}
$p(n_1,n_2,n_3)$ is the joint probability of 
having $n_1\mu$, $n_2\mu$ and $n_3\mu$ mutations 
respectively on the branches 
$(\alpha)$, $(\beta)$ and $((\alpha\beta)\,$$\cup$$\,(\gamma))$ 
of the tree in Fig.\,\ref{tree3}.
The factor 2 is an exchange factor which takes into 
account the possibility 
of having $n_1\,$$=$$\,n(\alpha)$ or equivalently
$n_1\,$$=$$\,n(\beta)$  if  
$n(\alpha)\,$$\neq$$\,n(\beta)$.

Since the three events are independent, using the exponential
distributions of the 
coalescent times $t_n$, i.e. of the length of the branches of the genealogical tree, we have:
\begin{equation}
p(n_1,n_2,n_3)=
2\int_0^{\infty}\int_0^{\infty}
p_\mu(n_1|t_3)\,p_\mu(n_2|t_3)\,
p_\mu(n_3|2t_2+t_3)\,3e^{-t_2+3\,t_3}dt_2\,dt_3\,,
\label{A20}
\end{equation}
where the factor 2 comes from the Jacobian of the
transformation 
$t'\,$$=$$\,t'(t_2,t_3)\,$$=$$\,2t_2\,$$+$$\,t_3$.
 
The conditional probabilities in the 
integral are all of the form $p_\mu(n|t)$ 
as given by the expression
\begin{equation}
p_\mu(n|t)=\left(\begin{array}{c}tN \\
n\mu\end{array}\right)\left(\frac{\mu}{2N}\right)^{n\cdot\mu}
\left(1-\frac{\mu}{2N}\right)^{tN-n\cdot\mu}\sim \frac{e^{-\mu t/2}(\mu t/2)^{n\mu}}{(n\mu)!}\,,
\label{pn}
\end{equation}
which is easily derivable from equation\,(\ref{kt})
and the second approximations holds for
large populations ($N \to \infty$).
  
In the same way one has that the probability 
$P(B)$ of event $B$ 
is the probability of having $n_1\,$$>$$\,n_2$ and $n_3\,$$=$$\,n_1$. 
Then we obtain:
\begin{equation}
P(B)=2\sum_{n_1=1/\mu}^{\infty}\sum_{n_2=0}^{n_1-1/\mu}p_{\mu}(n_1,n_2,n_1)\,
\end{equation}
where the factor 2 is also an exchange factor.

Finally, for the probability $P(C)$ of event $C$, that is the
probability of having 
$n_1\,$$=$$\,n_2\,$$=$$\,n_3$, we can write:
\begin{equation}
P(C)=\sum_{n_1=0}^{\infty}p_{\mu}(n_1,n_1,n_1)\,,
\end{equation}
where, of course, there is no exchange factor.

Putting together all the different terms, the resulting expression is: 
\begin{eqnarray}
P(R)&=&3\int_0^{\infty}\int_0^{\infty} P(R|t_2,t_3)e^{-(3t_3+t_2)}dt_2\,dt_3\,,
\end{eqnarray}
and 

\begin{eqnarray}
P(R|t_2,t_3)&=&\sum_{n_1=\frac{1}{\mu}}^{\infty}\sum_{n_2=0}^{n_1-
\frac{1}{\mu}}\sum_{n_3=n_1+\frac{1}{\mu}}^{\infty}2p_{\mu}(n_1|t_3)p_{\mu}(n_2|t_3)p_{\mu}(n_3|t_3+2t_2)+\nonumber\\
&+&\sum_{n_1=0}^{\infty}\sum_{n_3=n_1+\frac{1}{\mu}}^{\infty}p_{\mu}(n_1|t_3)^2p_{\mu}(n_3|t_3+2t_2)+\nonumber\\
&+&\sum_{n_1=\frac{1}{\mu}}^{\infty}\sum_{n_2=0}^{n_1-\frac{1}{\mu}}p_{\mu}(n_1|t_3)p_{\mu}(n_2|t_3)
p_{\mu}(n_1|t_3+2t_2)+\nonumber\\
&+&\sum_{n_1=0}^{\infty}\frac{1}{2}p_{\mu}(n_1|t_3)^2p_{\mu}(n_1|t_3+2t_2),
\end{eqnarray}

Finally, the probability of a wrong reconstruction 
is $P(W)=1-P(R)$ \,.

\newpage

\end{document}